\documentclass[aps,prl,twocolumn,superscriptaddress]{revtex4}
\usepackage{graphicx}
\usepackage{mathrsfs}

\begin{document}

\title{Doppler controlled dynamics of a mirror attached to a spring}

\author{K.Karrai}
\email{k.karrai@lmu.de} \affiliation{Center for Nanoscience and
Fakult\"{a}t f\"{u}r Physik, Ludwig-Maximilians-Universit\"{a}t,
Geschwister-Scholl-Platz 1, 80539 M\"{u}nchen, Germany\\}

\author{I.Favero}
\email{ivan.favero@physik.uni-muenchen.de} \affiliation{Center for
Nanoscience and Fakult\"{a}t f\"{u}r Physik,
Ludwig-Maximilians-Universit\"{a}t, Geschwister-Scholl-Platz 1,
80539 M\"{u}nchen, Germany\\}

\author{C.Metzger}
\affiliation{Center for Nanoscience and Fakult\"{a}t f\"{u}r
Physik, Ludwig-Maximilians-Universit\"{a}t,
Geschwister-Scholl-Platz 1, 80539 M\"{u}nchen, Germany\\}

\date{19 June 2007}

\begin{abstract}

A laser beam directed at a mirror attached onto a flexible mount
extracts thermal energy from its mechanical Brownian motion by
Doppler effect. For a normal mirror the efficiency of this Doppler
cooling is very weak and masked by laser shot-noise. We find that
it can become really efficient using a Bragg mirror at the long
wavelength edge of its band stop. The opposite effect opens new
routes for optical pumping of mechanical systems: a laser pointing
at a Bragg mirror and tuned at its short wavelength edge induces
amplification of the vibrational excitation of the mirror leading
eventually to its self-oscillation. This new effects rely on the
strong dependency of the Bragg mirror reflectivity on the
wavelength.

\end{abstract}


\maketitle

Radiation pressure and even electromagnetic fluctuation of the
vacuum are known to affect the dynamics of mirrors attached to a
spring \cite{a,b} but less is known about relativistic effects of
light on such macroscopic mechanical oscillators. Recent results
show that opto-mechanical couplings can be exploited in a
deformable Fabry-Perot arrangement to cool down or enhance the
mechanical fluctuations of spring loaded tiny mirrors \cite{c}.
This effect exploits the fact that inside a Fabry-Perot cavity the
photon-pressure reacts to any change in the mirror position with a
time delay given by the cavity photon storage time
\cite{c,d,e,f,g}. These cavity effects, and notably the cavity
laser cooling, have much analogy with Doppler cooling of atoms yet
they do not require relativistic effects for their interpretation.
In this work we describe a mechanism for true Doppler modified
dynamics of a mirror thermally and mechanically anchored to a
spring. Without the use of any cavity, this mechanism could allow
to set optically the mirror into a regime of mechanical
self-oscillation, simply by pointing a laser at it.

A laser beam at wavelength   and power P illuminates the mirror of
reflectivity R($\lambda$) as schematized in fig. 1. The radiation
pressure acting on a moving mirror varies in proportion to the
Doppler shift. We show here that because of this dynamics, the
Brownian fluctuation of the suspended mirror looses its energy to
the electromagnetic field. We found that the efficiency of Doppler
cooling can be much increased using a mirror with a large and
negative gradient of R($\lambda$). We also show that an
amplification of the vibrational excitation can be reached for
positive large enough gradient of R($\lambda$). Large gradient of
reflectivity are usually found at the edges of the band-stop of a
Bragg mirror.

\begin{figure}[t]
\includegraphics[width=9cm]{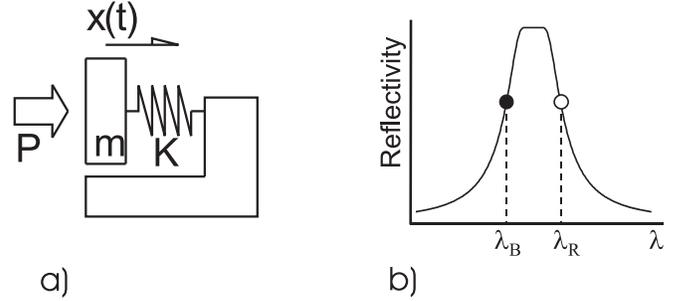}
\caption{(a) Schematics of a mirror of reflectivity R($\lambda$)
and mass m mechanically attached and thermally anchored to a
spring of rigidity K. The friction mechanisms internal to the
spring are responsible for losses of mechanical energy at a rate
$\Gamma$. A laser of power P is directed at the mirror from the
left. (b) Schematics of the reflectivity wavelength dependency of
a Bragg mirror. The gradients of R are maximized at the edges
$\lambda_{B}$ and $\lambda_{R}$ of the mirror band stop. }
\label{resonance_combs}
\end{figure}

The movable mirror of mass m is subject to a force Fph due to
radiation pressure. Its position x and velocity v of the center of
mass of the mirror obey the Newton's equation of motion, namely
m(dv/dt)+ m$\Gamma$v +Kx = $F_{ph}$ + $F_{th}$. The Langevin force
$F_{th}$ is introduced here in order to account for the Brownian
fluctuation of the center mass of the mirror coupled to a thermal
bath at temperature T. The damping rate $\Gamma$ of the mirror
mechanical fluctuation is a factor characteristic of the spring
holding the mirror. The momentum transferred to the mirror per
photon is given by $\hbar$($k_{0}$-$k_{R}$). Here the incoming
photons have their wave vector $k_{0}$ = 2$\pi$/$\lambda$ and the
reflected ones have $k_{R}$ oriented in the opposite direction. In
the reference frame of the laboratory, when the mirror moves away
from the light source, the reflected photons have their momentum
reduced by Doppler effect such $k_{R}$ = -$k_{0}$(1-2v/c). The
radiation pressure is given by the rate of photon momentum
transfer to the mirror and is also reduced by Doppler effect such
that $F_{ph}$ = RdN/dt $\hbar$($k_{0}$-$k_{R}$) where dN/dt is the
number of impinging photon per unit time. The dependency of the
radiation pressure on v is $F_{ph}$ =
2R(dN/dt)$\hbar$$k_{0}$(1-v/c) and by making use of the laser
power P = $\hbar$$k_{0}$c(dN/dt) it is also $F_{ph}$ =
(2RP/c)(1-v/c). In addition, when the mirror reflectivity is a
function of $\lambda$ as it is the case for a Bragg mirror
operated near its band-stop edge, R also depends on the mirror
velocity through the Doppler effect. In this case, we expand
R[$\lambda$(v/c)] in the experimentally relevant limit of v/c
$\ll$ 1. To the first order in k, the reflectivity is R(v/c)
$\simeq$ $R_{0}$+(v/c)$\lambda$(dR/d$\lambda$), so in the same
way, to the first order in v/c the radiation pressure is $F_{ph}$
$\simeq$
(2$R_{0}$P/c)[1-v/c+(v/c)($\lambda$/$R_{0}$)(dR/d$\lambda$)].
Using this expression in the equation of motion and grouping the
velocity terms together we obtain an effective Newton equation of
motion m(dv/dt) + m$\Gamma_{eff}$v + Kx = $F_{ph,0}$ + $F_{th}$
with a constant radiation pressure $F_{ph,0}$ = 2$R_{0}$P/c and a
Doppler modified damping rate

\begin{equation}
\Gamma_{eff}/\Gamma =
1+(2R_{0}P/mc^{2}\Gamma)[1-(\lambda/R_{0})(dR/d\lambda)]
\end{equation}

The constant force $F_{ph,0}$ only shifts the average position of
the center of mass and will be ignored in solving the effective
equation of motion. For dR/d$\lambda$ $\leq$ 0, the optical
contribution to the effective dissipation term $\Gamma_{eff}$
takes energy away from the mechanical Brownian fluctuation and
turns it irreversibly into electromagnetic energy through Doppler
effect. This amounts to cooling of the Brownian fluctuations of
the mirror. We now determine the temperature of the vibrational
motion of the mirror. For a harmonic oscillator the equipartition
theorem links the temperature to the time averaged amplitude
$\langle xx^{*}\rangle$ of the Brownian fluctuation, namely
(1/2)$k_{B}$$T_{eff}$ = (1/2)K$\langle xx^{*}\rangle$. In
experimental conditions it is not the temporal dependency $x(t)$
but rather its spectral distribution $x_{\omega}$ which is
typically measured. The spectrum $x_{\omega}$ is in fact a Fourier
transform of $x(t)$. A mathematically convenient property of
Fourier transformation is that the time averaged value $\langle
x(t)x^{*}(t)\rangle$ term equals the frequency averaged value
$\langle x_{\omega}x_{\omega}^{*}\rangle$. In Fourier space the
Newton's effective equation is
(-m$\omega^{2}$+im$\Gamma_{eff}$$\omega$+K)$x_{\omega}$ =
$F_{th,\omega}$, from which we obtain the spectrum $x_{\omega}$ =
($F_{th,\omega}$/m)/(-$\omega^{2}$+i$\Gamma_{eff}$$\omega$+K/m).
Here for a non-absorbing mirror, the spectral decomposition
$F_{th,\omega}$ of the thermal fluctuation driving force does not
depend on the light and is evaluated from the situation in dark. A
reasonable guess about the nature of the driving force
$F_{th,\omega}$ is that there is no preferred frequency for the
thermal fluctuations in the range of the mirror mechanical
vibrational frequencies. This is reasonable at vibrational
mechanical frequency much lower compared to typical phonon
frequencies with high density of phonon modes within the mirror
material (THz). Within this approximation we assume the spectral
power density $S_{th}$ in thermal excitation of the mirror to be
frequency independent, such that for any given frequency window d
the amplitude of the thermal driving force obeys
$F_{th,\omega}F^{^{*}}_{th,\omega}$ = $S_{th}$d$\omega$. Using the
expression of $x_{\omega}$ given above we obtain after some
algebra (1/2)K$\langle x_{\omega}x_{\omega}^{*}\rangle$ =
$\pi$$S_{th}$/(4m$\Gamma_{eff}$). The left hand side this equation
is (1/2)$k_{B}$T$_{eff}$ as prescribed by the equipartition
theorem and we end up with (1/2)$k_{B}$T$_{eff}$ =
$\pi$$S_{th}$/(4m$\Gamma_{eff}$). Since the spectral power density
$S_{th}$ is not light dependent, in dark we also have 1/2$k_{B}$T
= $\pi$$S_{th}$/(4m$\Gamma$). Comparing both expressions we obtain
T/T$_{eff}$=$\Gamma_{eff}$/$\Gamma$, showing that an increase in
$\Gamma_{eff}$ leads to cooling. This conclusion is premature
because so far we ignored the effect of photon shot noise. For a
given laser intensity we need to include the fundamental
shot-noise power fluctuation that induces a corresponding
shot-noise in the radiation pressure and hence driving an
additional vibrational fluctuation or an added vibrational
temperature. This added fluctuation could counteract the Doppler
cooling. This fluctuation force $F_{shot}$ is very much analogous
to $F_{th}$ but is proportional to square root of the laser power.
Here $F_{shot,\omega}$$F^{^{*}}_{shot,\omega}$ =
$S_{shot}$d$\omega$ with a frequency independent power density
$S_{shot}$=(2$R_{0}$/c)$^{2}$(Ph$\nu$)/2$\pi$ where h$\nu$ is the
photon energy. Assuming no correlation between the Brownian and
shot noise, we obtain the new prescription (1/2)KT$_{eff}$ =
$\pi$($S_{th}+S_{shot}$)/(4m$\Gamma_{eff}$) leading to the
expression for the effective vibrational temperature:

\begin{equation}
T_{eff} = (T+pT_{phot})/(1+p(1-\nabla R))
\end{equation}

where we defined i) the unitless reduced laser power p =
2$R_{0}$P/(m$c^{2}$$\Gamma$), ii) the reflectivity gradient
$\nabla$R = ($\lambda$/$R_{0}$)(dR/d$\lambda$) and iii) an
effective photon temperature $k_{B}$T$_{phot}$ = $R_{0}$h$\nu$/2.
The power dependent term in the denominator originates from the
Doppler effect while the one in the numerator is due to the photon
shot-noise. We see that for a mirror with a constant reflectivity
dR/d$\lambda$ = 0, the Doppler effect tends to lower the effective
temperature while the shot noise terms increases it. For such a
mirror, the condition T$_{eff}$ $<$ T, namely for cooling is only
possible when T$_{phot}$ $<$ T or $R_{0}$h$\nu$/2 $<$ $k_{B}$T.
For visible or near-infrared photons and for high reflectivity
mirrors $R_{0}$ $\sim$ 1, this condition cannot be satisfied at
room temperature. For a mirror with a large and negative
reflectivity gradient however, such that $\nabla$R $<$ 0, Doppler
cooling of the vibrational mode becomes possible when
T$_{phot}$/(1-$\nabla$R) $<$ T or $R_{0}$h$\nu$/[2(1-$\nabla$R)]
$<$ $k_{B}$T. This gives a stringent condition for the
reflectivity gradient. At room temperature using 1 eV photons on a
mirror with $R_{0}$ $\sim$ 0.5, namely for T$_{phot}$ = 2900 K,
the condition on the reflectivity gradient would be -$\nabla$R $>$
8.7. Experimentally this can be obtained using a Bragg reflector
and a photon wavelength tuned at the higher wavelength-edge of the
band-stop of the reflectivity (fig. 1b). Values as high and
negative as ($\lambda$/$R_{0}$)(dR/d$\lambda$) = -$10^{5}$ are in
fact within experimental reach in the visible and near infrared
\cite{i}. Doppler cooling saturates using large enough laser power
at T$_{min}$ = T$_{phot}$/(1-$\nabla$R). With the numerical
example above this would be 29 mK.

Another interesting aspect of Bragg mirrors is that they offer
also the opportunity to set the gradient $\nabla$R positive enough
to reach  $\Gamma_{eff} \leq 0$ in equation (1). In this regime
the mirror gains energy and starts to self-oscillate under the
illumination of a CW laser. It enters possibly in a regime of
nonlinear dynamics similar to the one predicted by F. Marquardt
and coworkers \cite{g} for deformable Fabry-Perot cavities. This
effect would be interesting to demonstrate in this context of
cavity-less system. The gain condition implies using a laser power
large enough so that P/$\Gamma$ $>$
m$c^{2}$/[2$R_{0}$($\nabla$R-1)]. The energy term P/$\Gamma$
represents the laser power averaged over a mechanical relaxation
time constant 1/$\Gamma$. This energy is compared to m$c^{2}$, the
relativistic energy of the mirror so we anticipate already the
Doppler induced optomechanic might be very weak for macroscopic
mirrors. We recently prepared a gold mirror (dR/d$\lambda$ $\sim$
0) mounted on silicon nano-lever \cite{h} with a mass in the
$10^{-15}$ Kg range. For such a mirror and for $R_{0}$ $\sim$ 0.5,
the condition for self-oscillation would be reached when
P/$\Gamma$ $>$ 90/($\nabla$R-1). Using a damping rate $\Gamma$
$\sim$ 10 sec$^{-1}$ and $\nabla$R $\sim$ $10^{5}$ this would
imply using laser power of 90 mW in order to enter a regime of
self-oscillation. In order to increase the Doppler effect we see
from eq. (2) that one needs to decrease the mass m. For a Bragg
mirror, this can only be done within bounds because photons probe
the material periodicity within a finite penetration depth. The
mass cannot arbitrarily reduced by thinning the material, at some
point the reflectivity and its gradient will degrade. Also the
lateral dimensions of the reflector cannot be reduced much less
than the diffraction limit [8]. In the visible range we anticipate
that the smallest masses would be in the $10^{-15}$ Kg range.

Now we compare the strength of Doppler-cooling with cavity cooling
established in ref \cite{c,d,e,f}. A Fabry-Perot cavity of length
L separating the mirrors and finesse F$>$1 stores electromagnetic
energy with a typical ring-down time-constant $\tau$ $\sim$
(F/$\pi$)$\tau_{0}$ where $\tau_{0}$ = L/c is the photon time of
flight across the cavity. $\tau$ is the typical time the cavity
needs to build or loose energy upon a sudden change in laser power
or in mirror separation. The photon pressure acting on the mirrors
is proportional to the stored power so that the force acting on
the mirror near a cavity resonance is not only enhanced by the
cavity but also retarded with respect to the mirror separation
fluctuation. Retarded terms amounts to velocity dependent force
terms similarly to the Doppler effect discussed above. For a
cavity with at least one of the two mirrors mounted on a spring,
this retarded effect induces an optical modification of the
mechanical damping rate and consequently a modification of the
vibrational temperature the same way developed above. In order to
cool the vibrational fluctuations of the mirror, it is necessary
to detune slightly the laser wavelength to the red with respect to
the cavity resonance \cite{c,d,e,f}. Optimum detuning
$\delta$$\{L/\lambda\}$ is obtained on the maximum slope of the
dependency of electromagnetic energy stored in the cavity with
respect to wavelength or mirror separation. This is the case when
$\delta$$\{L/\lambda\}$ = -1/(g$\sqrt{3}$). The reverse effect,
namely opto-mechanical excitation, is obtained for blue detuning.
We establish the extremal effective temperatures

\begin{equation}
T_{eff} \simeq (T+pgT_{phot})/[1\pm
p(L/\lambda)g^{3}/(g^{2}\omega_{0}^{2}\tau_{0}^{2}+1)]
\end{equation}

Where g = 2F/$\pi$ is proportional to the cavity finesse F. The
sign depends on the side of the detuning with respect to a cavity
resonance. For a given vibrational frequency  $\omega_{0}$ =
$(K/m)^{1/2}$ it turns out that optimal cooling (or pumping) is
obtained at the condition $\omega_{0}\tau$ $\sim$ 1.

In expressing equation (3) we made the reduced power p =
2$R_{0}$P/m$c^{2}$$\Gamma$ appear explicitly in order to make a
direct comparison with Doppler cooling. The term in the
denominator is usually much larger than unity and we see that
cooling efficiency using cavity effects can be made stronger by a
factor as large as g$^{3}$(L/$\lambda$) than Doppler cooling.
Already for a finesse as low as g = 10 and for a length L = 20 000
$\lambda$, the cavity cooling can be made $10^{7}$ more efficient
than direct Doppler cooling and this considering equal mirror
masses, laser powers and damping rates. The use of large finesses
allows a significant amplification of laser cooling but at the
same time imposes that the moving mirror be part of a Fabry-Perot
cavity and this is not always convenient. We stress however the
fact that all models derived so far describing cavity cooling do
not include relativistic effects. While mimicking it, cavity
cooling cannot be interpreted as Doppler cooling. For completeness
we introduced Doppler effect in our formalism of cavity cooling
and found that it gives rise to corrections to the cooling
efficiency that are small for cavity finesses g $\gg$ 1 \cite{j}.

We finish this letter on an estimation of the cooling power
involved with Doppler effect.

In dark, the mechanical fluctuation dissipates its thermal energy
$k_{B}$T/2 per mechanical degree of freedom and this at a rate
$\Gamma$. The dissipated power is there-fore ($k_{B}$T/2)$\Gamma$
and is in equilibrium with the power that feeds the fluctuation.
When the mirror reflects the laser light, the effective
vibrational mode end-temperature is T$_{eff}$. When the
vibrational mode is cooled down to a temperature T$_{eff}$, the
steady state heat-load in the mirror is
($k_{B}$T$_{eff}$/2)$\Gamma$. Consequently in order to maintain a
temperature T$_{eff}$, the optical cooling extracts energy from
the fluctuations of the mirror position at a rate P$_{cooling}$ =
$k_{B}$(T-T$_{eff}$)$\Gamma$/2 which is always smaller than
$k_{B}$T$\Gamma$/2. So the maximum cooling power is
$k_{B}$T$\Gamma$/2 both for Doppler and cavity cooling. This is in
the range of $10^{-18}$ Watts at room temperature for $\Gamma$ in
the $10^{3}$ sec$^{-1}$ range. This might appear as a very weak
cooling power but it can be efficient enough to cool the lowest
energy vibrational modes of an elastically suspended mirror since
such modes are generally weakly coupled to the thermal bath.

In conclusion, we presented a simple formalism for laser Doppler
cooling of the center mass fluctuation of a mirror attached to a
spring. This effect is very weak but can become sizeable when the
mirror reflectivity is made to depend strongly on the photon
wavelength. We also showed that effective temperature obtained
through cavity cooling, in a formalism that does not include
Doppler effect, mimics direct Doppler cooling but with a cavity
amplification factor which is proportional to the third power of
the cavity finesse and can easily reach ten orders of magnitudes.
The reciprocal effect of Doppler cooling, namely Doppler optical
pumping of the mirror motion, was also predicted. Interestingly
enough, we showed that the use of an appropriate Bragg mirror
should allow overcoming the shot noise limit and use directly the
Doppler effect to set the mirror motion into self-oscillation.
This could provide a very simple and non-invasive method to
optically pump the motion of tiny mechanical resonators.

We thank Joel Chevrier, Yann Tissot and Hans Limberger for
stimulating discussions. The DFG grant NOMS KA 1216/2-1 supported
the contribution of C. Metzger. The Alexander Von Humboldt
foundation supported I. Favero.


\begin{references}



\bibitem{a} V. B. Braginsky, A. B. Manukin, {\it Measurements of weak forces in Physics experiments}, University of Chicago Press,
(1977).



\bibitem{b} H. B. Chan, V. A. Aksyuk, R. N. Kleiman, D. J. Bishop,
Federico Capasso, Science {\bf 291}, 1941 (2001).


\bibitem{c} C. Metzger and K. Karrai, Nature {\bf 432}, 1002 (2004).



\bibitem{d} O. Arcizet,  P. F. Cohadon, T. Briant, M. Pinard, A. Heidmann,
Nature {\bf 444}, 71 (2006).



\bibitem{e} S. Gigan, H. R. B\"{o}hm, M. Paternostro, F. Blaser, G. Langer,
J. B. Hertzberg, K. C. Schwab, D. B\"{a}uerle, M. Aspelmeyer, A.
Zeilinger, Nature {\bf 444}, 67 (2006).



\bibitem{f} A. Schliesser, P. Del'Haye, N. Nooshi, K. J. Vahala, T. J.
Kippenberg, Phys. Rev. Lett. {\bf 97}, 243905 (2006).



\bibitem{g} F. Marquardt, J. G. E. Harris, S. M. Girvin, Phys. Rev. Lett.
{\bf 96}, 103901 (2006).



\bibitem{h} I. Favero, C. Metzger, S. Camerer, D. K\"{o}nig, H. Lorenz, J. P.
Kotthaus, K. Karrai, Appl. Phys. Lett. {\bf 90}, 041101 (2007).



\bibitem{i} Y. Tissot, H. Limberger, private communication.



\bibitem{j} C. Metzger, I. Favero, K. Karrai, unpublished.



\end{references}
\end{document}